\documentstyle[12pt,epsf]{article}
\newcommand{\agt}{\,\rlap{\lower 3.5 pt \hbox{$\mathchar \sim$}} \raise 1pt
 \hbox {$>$}\,}
\newcommand{\alt}{\,\rlap{\lower 3.5 pt \hbox{$\mathchar \sim$}} \raise 1pt
 \hbox {$<$}\,}

\begin{document}
\title{\vskip-3cm{\baselineskip14pt
\centerline{\normalsize MPI/PhT/97--025\hfill}
\centerline{\normalsize hep--ph/9706430\hfill}
\centerline{\normalsize April 1997\hfill}}
\vskip1.5cm
Strong Coupling Constant with Flavour Thresholds at Four Loops in the
$\overline{\rm MS}$ Scheme}
\author{{\sc K.G. Chetyrkin}\thanks{Permanent address:
Institute for Nuclear Research, Russian Academy of Sciences,
60th October Anniversary Prospect 7a, Moscow 117312, Russia.}
{\sc, B.A. Kniehl, and M. Steinhauser}\\
Max-Planck-Institut f\"ur Physik (Werner-Heisenberg-Institut),\\
F\"ohringer Ring 6, 80805 Munich, Germany}
\date{}
\maketitle
\begin{abstract}
We present in analytic form the matching conditions for the strong coupling
constant $\alpha_s^{(n_f)}(\mu)$ at the flavour thresholds to three loops in
the modified minimal-subtraction scheme.
Taking into account the recently calculated coefficient $\beta_3$ of the
Callan-Symanzik beta function of quantum chromodynamics, we thus derive a
four-loop formula for $\alpha_s^{(n_f)}(\mu)$ together with appropriate
relationships between the asymptotic scale parameters $\Lambda^{(n_f)}$ for
different numbers of flavours $n_f$.

\end{abstract}
\newpage

The strong coupling constant $\alpha_s^{(n_f)}(\mu)=g_s^2/(4\pi)$, where $g_s$
is the gauge coupling of quantum chromodynamics (QCD), is a fundamental
parameter of the standard model of elementary particle physics;
its value $\alpha_s^{(5)}(M_Z)$ is listed among the constants of nature
in the Review of Particle Physics \cite{pdg}.
Here, $\mu$ is the renormalization scale, and $n_f$ is the number of active
quark flavours $q$, with mass $m_q\ll\mu$.
The $\mu$ dependence of $\alpha_s^{(n_f)}(\mu)$ is controlled by the
Callan-Symanzik beta function of QCD,
\begin{eqnarray}
\label{rge}
\frac{\mu^2d}{d\mu^2}\,\frac{\alpha_s^{(n_f)}(\mu)}{\pi}
&=&\beta^{(n_f)}\left(\frac{\alpha_s^{(n_f)}(\mu)}{\pi}\right)
\nonumber\\
&=&{}-\sum_{N=0}^\infty\beta_N^{(n_f)}
\left(\frac{\alpha_s^{(n_f)}(\mu)}{\pi}\right)^{N+2}.
\end{eqnarray}
The calculation of the one-loop coefficient $\beta_0^{(n_f)}$ about 25 years
ago \cite{gro} has led to the discovery of asymptotic freedom and to the
establishment of QCD as the theory of strong interactions.
In the class of schemes where the beta function is mass independent, which
includes the minimal subtraction (MS) schemes of dimensional regularization
\cite{bol}, $\beta_0^{(n_f)}$ and $\beta_1^{(n_f)}$ \cite{jon} are universal.
The results for $\beta_2^{(n_f)}$ \cite{tar} and $\beta_3^{(n_f)}$ \cite{rit}
are available in the modified MS ($\overline{\rm MS}$) scheme \cite{bar}.
For the reader's convenience, $\beta_N^{(n_f)}$ $(N=0,\ldots,3)$ are listed
for the $n_f$ values of practical interest in Table~\ref{t1}.

\begin{table}[h]
\begin{center}
\caption{$\overline{\rm MS}$ values of $\beta_N^{(n_f)}$ for variable $n_f$.}
\label{t1}
\begin{tabular}{c|cccc}
\hline
\hline
$n_f$ & $\beta_0^{(n_f)}$ & $\beta_1^{(n_f)}$ & $\beta_2^{(n_f)}$ &
$\beta_3^{(n_f)}$ \\
\hline
$3$
& $\frac{9}{4}$
& $4$
& $\frac{3863}{384}$
& $\frac{445}{32}\zeta(3)+\frac{140599}{4608}$
\\
$4$
& $\frac{25}{12}$
& $\frac{77}{24}$
& $\frac{21943}{3456}$
& $\frac{78535}{5184}\zeta(3)+\frac{4918247}{373248}$
\\
$5$
& $\frac{23}{12}$
& $\frac{29}{12}$
& $\frac{9769}{3456}$
& $\frac{11027}{648}\zeta(3)-\frac{598391}{373248}$
\\
$6$
& $\frac{7}{4}$
& $\frac{13}{8}$
& $-\frac{65}{128}$
& $\frac{11237}{576}\zeta(3)-\frac{63559}{4608}$
\\
\hline
\hline
\end{tabular}
\end{center}
\end{table}

In MS-like renormalization schemes, the Appelquist-Carazzone decoupling
theorem \cite{app} does not in general apply to quantities that do not
represent physical observables, such as beta functions or coupling constants, 
{\it i.e.}, quarks with mass $m_q\gg\mu$ do not automatically decouple.
The standard procedure to circumvent this problem is to render decoupling
explicit by using the language of effective field theory.
As an idealized situation, consider QCD with $n_l=n_f-1$ massless quark
flavours and one heavy flavour $h$, with mass $m_h\gg\mu$.
Then, one constructs an effective $n_l$-flavour theory by requiring 
consistency with the full $n_f$-flavour theory at the heavy-quark threshold
$\mu^{(n_f)}={\cal O}(m_h)$.
This leads to a nontrivial matching condition between the couplings of the two
theories.
Although, $\alpha_s^{(n_l)}(m_h)=\alpha_s^{(n_f)}(m_h)$ at leading and
next-to-leading order, this relation does not generally hold at higher orders
in the $\overline{\rm MS}$ scheme.
If the $\mu$ evolution of $\alpha_s^{(n_f)}(\mu)$ is to be performed at $N+1$
loops, {\it i.e.}, with the highest coefficient in Eq.~(\ref{rge}) being
$\beta_N^{(n_f)}$, then consistency requires that the matching conditions be
implemented in terms of $N$-loop formulae.
Then, the residual $\mu$ dependence of physical observables will be of order
$N+2$.
A pedagogical review of the QCD matching conditions at thresholds to two loops
may be found in Ref.~\cite{rod}.

The literature contains two conflicting results on the two-loop matching
condition for $\alpha_s^{(n_f)}(\mu)$ in the $\overline{\rm MS}$ scheme
\cite{ber,lar}.
The purpose of this letter is to settle this issue by an independent 
calculation and to take the next step, to three loops.
As a consequence, Eq.~(9.7) in the encyclopedia by the Particle Data Group
\cite{pdg} will be corrected and extended by one order.
We shall also add the four-loop term in the formula~(9.5a) for
$\alpha_s^{(n_f)}(\mu)$ in Ref.~\cite{pdg}.

In order to simplify the notation, we introduce the couplant
$a^{(n_f)}(\mu)=\alpha_s^{(n_f)}(\mu)/\pi$ and omit the labels $\mu$ and $n_f$
wherever confusion is impossible.
Integrating Eq.~(\ref{rge}) leads to
\begin{eqnarray}
\label{con}
\ln\frac{\mu^2}{\Lambda^2}&=&\int\frac{da}{\beta(a)}
\nonumber\\
&=&\frac{1}{\beta_0}\left[\frac{1}{a}+b_1\ln a+(b_2-b_1^2)a
\right.\nonumber\\
&+&\left.
\left(\frac{b_3}{2}-b_1b_2+\frac{b_1^3}{2}\right)a^2\right]+C,
\end{eqnarray}
where $b_N=\beta_N/\beta_0$ ($N=1,2,3$), $\Lambda$ is the so-called asymptotic
scale parameter, and $C$ is an arbitrary constant.
The second equation of Eq.~(\ref{con}) is obtained by expanding the integrand.
The conventional $\overline{\rm MS}$ definition of $\Lambda$, which we shall
adopt in the following, corresponds to choosing $C=(b_1/\beta_0)\ln\beta_0$
\cite{bar,fur}.

Iteratively solving Eq.~(\ref{con}) yields
\begin{eqnarray}
\label{alp}
a&=&\frac{1}{\beta_0L}-\frac{b_1\ln L}{(\beta_0L)^2}
+\frac{1}{(\beta_0L)^3}\left[b_1^2(\ln^2L-\ln L-1)+b_2\right]
\nonumber\\
&+&\frac{1}{(\beta_0L)^4}\left[
b_1^3\left(-\ln^3L+\frac{5}{2}\ln^2L+2\ln L-\frac{1}{2}\right)
\right.\nonumber\\
&-&\left.
3b_1b_2\ln L+\frac{b_3}{2}\right],
\end{eqnarray}
where $L=\ln(\mu^2/\Lambda^2)$ and terms of ${\cal O}(1/L^5)$ have been
neglected.
Equation~(\ref{alp}) extends Eq.~(9.5a) of Ref.~\cite{pdg} to four loops.

The particular choice of $C$ \cite{bar,fur} in Eq.~(\ref{con}) is predicated
on the grounds that it suppresses the appearance of a term proportional to
$({\rm const.}/L^2)$ in Eq.~(\ref{alp}).
For practical applications, it might be more useful to define $C$ by equating
the one- and two-loop expressions of $\alpha_s(\mu)$, {\it i.e.}, by
nullifying the ${\cal O}(1/L^2)$ term in Eq.~(\ref{alp}), at some convenient
reference scale $\mu_0$ \cite{abb}, {\it e.g.}, at $\mu_0=M_Z$.
By contrast, in the standard approach, one has $\mu_0=\sqrt e\Lambda$, which
is in the nonperturbative regime.
This would lead to the choice
$C=(b_1/\beta_0)\ln[\beta_0\ln(\mu_0^2/\Lambda^2)]$.
The advantage of this convention would be that the values of $\Lambda$ would
be considerably more stable under the inclusion of higher-order corrections.
Another interesting alternative is to adjust $C$ in such a way that $\Lambda$ 
becomes $n_f$ independent \cite{mar}.

\begin{figure}[ht]
\begin{center}
\epsfxsize=14cm
\epsffile[106 275 469 571]{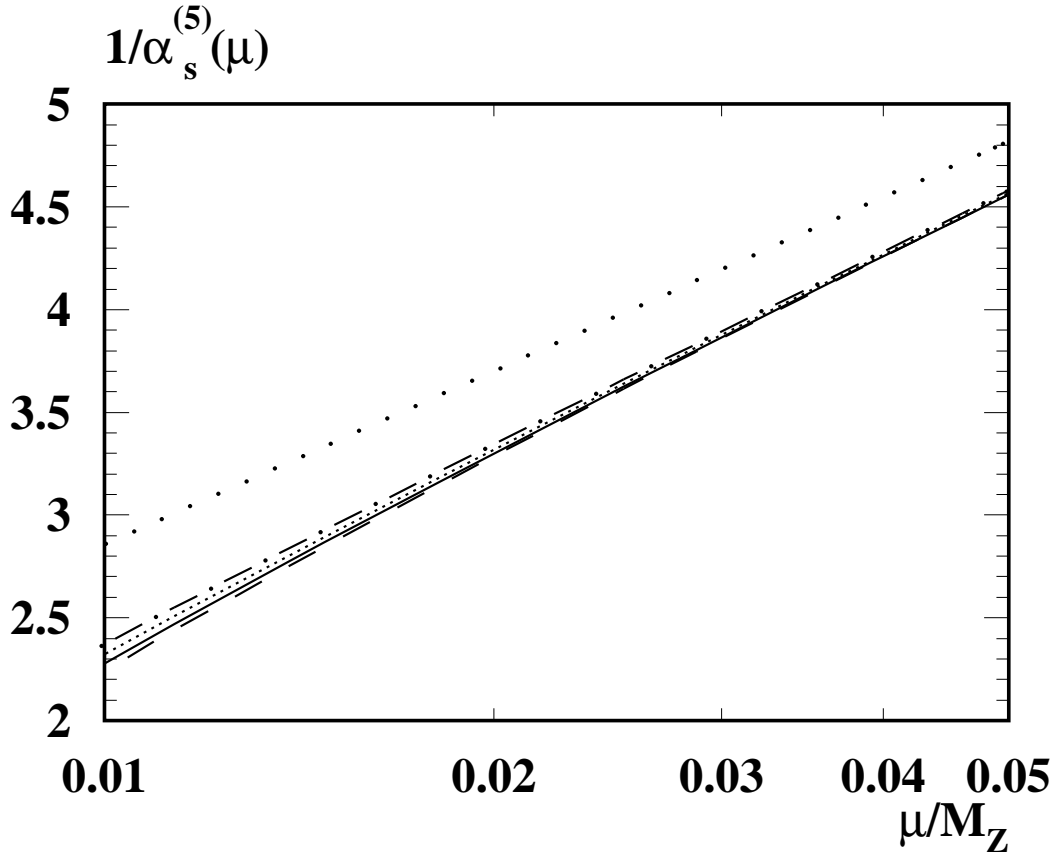}
\smallskip
\caption{$\mu$ dependence of $\alpha_s^{(5)}(\mu)$ calculated from
$\alpha_s^{(5)}(M_Z)=0.118$ using Eq.~(\ref{alp}) at one (coarsely dotted),
two (dashed), three (dot-dashed), and four (solid) loops.
The densely dotted line represents the exact solution of Eq.~(\ref{rge}) at 
four loops.}
\label{f1}
\end{center}
\end{figure}

It is interesting to quantitatively investigate the impact of the higher-order
terms of the beta function in Eq.~(\ref{rge}) on the $\mu$ dependence of
$\alpha_s^{(n_f)}$ for fixed $n_f$.
For illustration, we consider, as an extreme case, the evolution of
$\alpha_s^{(5)}(\mu)$ from $\mu=M_Z$ down to scales of the order of the proton
mass.
Specifically, we employ the four-loop formula~(\ref{alp}) and its $N$-loop
approximations, with $N=1,2,3$, which emerge from Eq.~(\ref{alp}) by
discarding the terms of ${\cal O}(1/L^{N+1})$.
In each case, we determine $\Lambda^{(5)}$ from the condition that
$\alpha_s^{(5)}(M_Z)=0.118$ \cite{pdg} be exactly satisfied.
For comparison, we also consider the exact solution of Eq.~(\ref{rge}) with
all known beta-function coefficients included.
In Fig.~\ref{f1}, the various results for $1/\alpha_s^{(5)}(\mu)$ are plotted 
versus $\mu/M_Z$ using a logarithmic scale on the abscissa.
Consequently, the one-loop result appears as a straight line.
All curves precisely cross at $\mu=M_Z$, outside the figure.
We observe that, for $N$ increasing, the expanded $N$-loop results of
Eq.~(\ref{alp}) gradually approach the exact four-loop solution of
Eq.~(\ref{rge}) in an alternating manner.
Down to rather low scales, the two-loop result already provides a remarkably
useful approximation to the exact four-loop result, while the one-loop result
is far off.

\begin{figure}[ht]
\begin{center}
\epsfxsize=14cm
\epsffile[136 634 470 722]{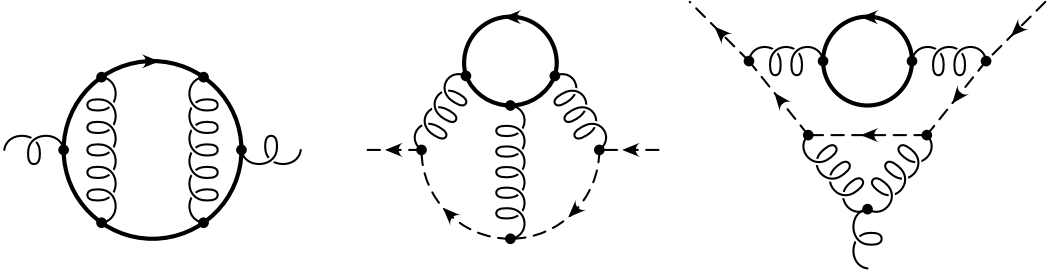}
\smallskip
\caption{Typical three-loop diagrams pertinent to $\Pi_G^h(q_G^2)$,
$\Pi_c^h(q_c^2)$, and $\Gamma_\mu^h(q_c,q_{\bar c})$.
Loopy, dashed, and solid lines represent gluons $G$, Faddeev-Popov ghosts $c$,
and heavy quarks $h$, respectively.}
\label{f2}
\end{center}
\end{figure}

Next, we outline the derivation of the three-loop matching condition.
In the following, unprimed quantities refer to the full $n_f$-flavour theory, 
while primed objects belong to the effective theory with $n_l=n_f-1$ flavours.
Futhermore, bare quantities are labelled by the superscript 0.
We wish to derive the decoupling constant $\zeta_g$ in the relation
$g_s^\prime=\zeta_gg_s$ between the renormalized couplings $g_s$ and
$g_s^\prime$.
Exploiting knowledge \cite{rit} of the coupling renormalization constant $Z_g$
within either theory, this task is reduced to finding
$\zeta_g^0=\zeta_gZ_g^\prime/Z_g$.
The Ward identity
$\zeta_g^0=\tilde\zeta_1^0/\left(\tilde\zeta_3^0\sqrt{\zeta_3^0}\right)$,
where
\begin{equation}
G_\mu^{0\prime}=\sqrt{\zeta_3^0}G_\mu^{0},\quad
c^{0\prime}=\sqrt{\tilde\zeta_3^0}c^0,\quad
\Lambda_\mu^{0\prime}=\tilde\zeta_1^0\tilde\zeta_3^0\sqrt{\zeta_3^0}
\Lambda_\mu^0,
\end{equation}
with $G_{a\mu}$, $c$, and $\Lambda_\mu$ being the fields of the gluon and the
Faddeev-Popov ghost, and the $G\bar cc$ vertex, respectively, then leads us to
consider the heavy-quark contributions to the corresponding vacuum
polarizations and vertex correction, $\Pi_G^h(q_G^2)$, $\Pi_c^h(q_c^2)$, and
$\Gamma_\mu^h(q_c,q_{\bar c})$.
Specifically, we have
\begin{eqnarray}
\zeta_3^0&=&1+\Pi_G^{0h}(0),\quad
\tilde\zeta_3^0=1+\Pi_c^{0h}(0),
\nonumber\\
\tilde\zeta_1^0&=&1+\left.\frac{q^\mu\Gamma_\mu^{0h}(q,-q)}{q^2}\right|_{q=0}.
\end{eqnarray}
In total, we need to compute $3+1+5$ two-loop and $189+25+228$ three-loop
Feynman diagrams.
The 5 two-loop diagrams pertinent to $\tilde\zeta_1^0$ add up to zero.
Typical three-loop specimen are depicted in Fig.~\ref{f2}.
In order to cope with the enormous complexity of the problem at hand, we make
successive use of powerful symbolic manipulation programs.
We generate and compute the relevant diagrams with the packages QGRAF
\cite{nog} and MATAD \cite{ste}, respectively.
The cancellation of the ultraviolet singularities, the gauge-parameter
independence, and the renormalization-group (RG) invariance serve as strong
checks for our calculation.

If we measure the matching scale $\mu^{(n_f)}$ in units of the RG-invariant
$\overline{\rm MS}$ mass $\mu_h=m_h(\mu_h)$, our result for the ratio of
$a^\prime=a^{(n_l)}(\mu^{(n_f)})$ to $a=a^{(n_f)}(\mu^{(n_f)})$ reads
\begin{eqnarray}
\label{msb}
\frac{a^\prime}{a}&=&1-a\frac{\ell_h}{6}
+a^2\left(\frac{\ell_h^2}{36}-\frac{19}{24}\ell_h+c_2\right)
+a^3\left[-\frac{\ell_h^3}{216}
\right.\nonumber\\
&-&\left.\vphantom{\frac{\ell_h^3}{216}}
\frac{131}{576}\ell_h^2+\frac{\ell_h}{1728}(-6793+281\,n_l)+c_3\right],
\end{eqnarray}
where $\ell_h=\ln[(\mu^{(n_f)})^2/\mu_h^2]$ and
\begin{equation}
\label{cms}
c_2=\frac{11}{72},\quad
c_3=-\frac{82043}{27648}\zeta(3)+\frac{564731}{124416}
-\frac{2633}{31104}n_l.
\end{equation}
Here, $\zeta$ is Riemann's zeta function, with values $\zeta(2)=\pi^2/6$ and
$\zeta(3)\approx1.202\,057$.
Our result for $c_2$ agrees with Ref.~\cite{lar}, while it disagrees with 
Ref.~\cite{ber}.
For the convenience of those readers who prefer to deal with the pole mass
$M_h$, we list here a simple formula \cite{kni} for $\mu_h$ in terms of $M_h$
and $A=a^{(n_f)}(M_h)$, which incorporates the well-known two-loop relation
between $m_h(M_h)$ and $M_h$ \cite{gra}.
It reads
\begin{eqnarray}
\frac{\mu_h}{M_h}&=&1-\frac{4}{3}A+A^2\left\{\frac{\zeta(3)}{6}
-\frac{\zeta(2)}{3}(2\ln2+7)-\frac{2393}{288}
\right.\nonumber\\
&+&\left.\frac{n_f}{3}\left[\zeta(2)+\frac{71}{48}\right]
\right\}.
\end{eqnarray}
Using a similar relation, with $A$ expressed in terms of $a$ and
${\cal L}_h=\ln[(\mu^{(n_f)})^2/M_h^2]$, we may rewrite Eq.~(\ref{msb}) as
\begin{eqnarray}
\label{oms}
\frac{a^\prime}{a}&=&1-a\frac{{\cal L}_h}{6}
+a^2\left(\frac{{\cal L}_h^2}{36}-\frac{19}{24}{\cal L}_h+C_2\right)
+a^3\left[-\frac{{\cal L}_h^3}{216}
\right.\nonumber\\
&-&\left.\vphantom{\frac{{\cal L}_h^3}{216}}
\frac{131}{576}{\cal L}_h^2+\frac{{\cal L}_h}{1728}(-8521+409\,n_l)+C_3
\right],
\end{eqnarray}
where
\begin{eqnarray}
C_2=-\frac{7}{24},\quad
C_3&=&-\frac{80507}{27648}\zeta(3)
-\frac{2}{3}\zeta(2)\left(\frac{1}{3}\ln2+1\right)
\nonumber\\
&-&\frac{58933}{124416}+\frac{n_l}{9}\left[\zeta(2)+\frac{2479}{3456}\right].
\end{eqnarray}

\begin{figure}[ht]
\begin{center}
\epsfxsize=14cm
\epsffile[90 275 463 579]{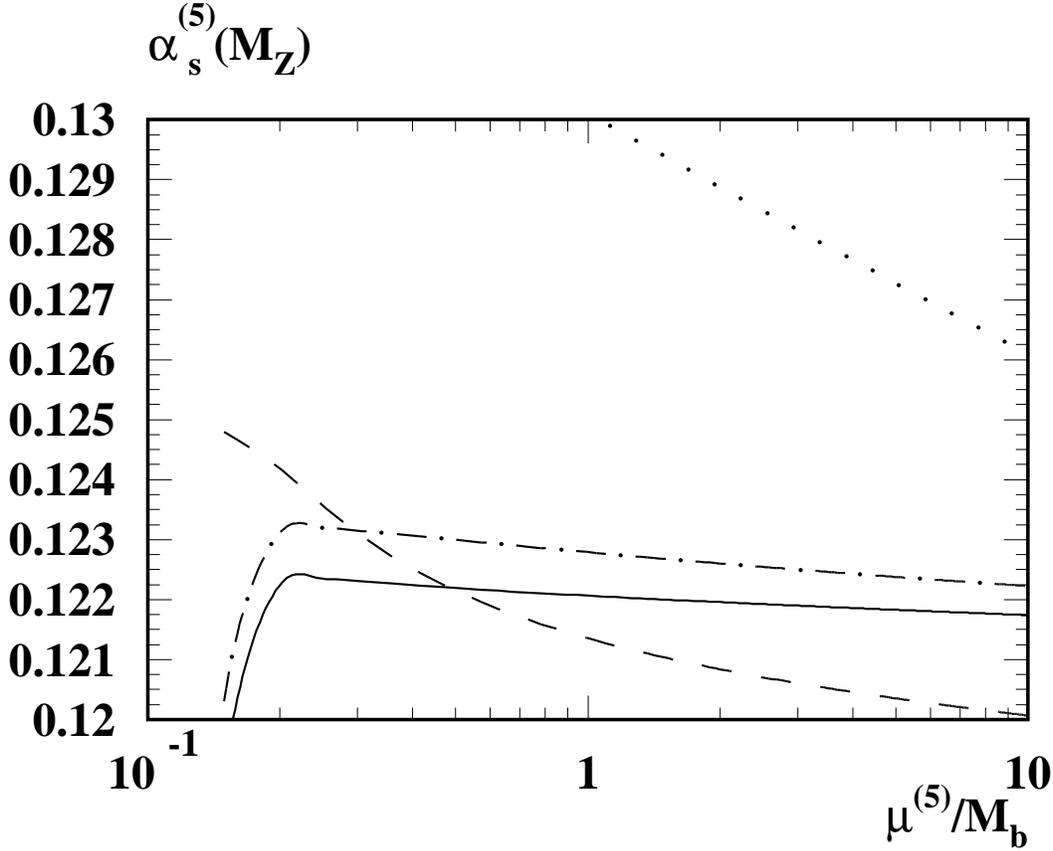}
\smallskip
\caption{$\mu^{(5)}$ dependence of $\alpha_s^{(5)}(M_Z)$ calculated from
$\alpha_s^{(4)}(M_\tau)=0.36$ and $M_b=4.7$~GeV using Eq.~(\ref{alp}) at one
(dotted), two (dashed), three (dot-dashed), and four (solid) loops in
connection with Eq.~(\ref{oms}) at the respective order.}
\label{f3}
\end{center}
\end{figure}

Going to higher orders, one expects, on general grounds, that the relation
between $\alpha_s^{(n_l)}(\mu^\prime)$ and $\alpha_s^{(n_f)}(\mu)$, where
$\mu^\prime\ll\mu^{(n_f)}\ll\mu$, becomes insensitive to the choice of
$\mu^{(n_f)}$ as long as $\mu^{(n_f)}={\cal O}(m_h)$.
This has been checked in Ref.~\cite{rod} for three-loop evolution in
connection with two-loop matching.
Armed with our new results, we are in a position to explore the situation at 
the next order.
As an example, we consider the crossing of the bottom-quark threshold.
In particular, we wish to study how the $\mu^{(5)}$ dependence of the relation
between $\alpha_s^{(4)}(M_\tau)$ and $\alpha_s^{(5)}(M_Z)$ is reduced as we
implement four-loop evolution with three-loop matching.
Our procedure is as follows.
We first calculate $\alpha_s^{(4)}(\mu^{(5)})$ with Eq.~(\ref{alp}) by
imposing the condition $\alpha_s^{(4)}(M_\tau)=0.36$ \cite{rod}, then obtain
$\alpha_s^{(5)}(\mu^{(5)})$ from Eq.~(\ref{oms}), and finally compute
$\alpha_s^{(5)}(M_Z)$ with Eq.~(\ref{alp}).
For consistency, $N$-loop evolution must be accompanied by $(N-1)$-loop 
matching, {\it i.e.}, if we omit terms of ${\cal O}(1/L^{N+1})$ in
Eq.~(\ref{alp}), we need to discard those of ${\cal O}(a^N)$ in
Eq.~(\ref{oms}) at the same time.
In Fig.~\ref{f3}, the variation of $\alpha_s^{(5)}(M_Z)$ with $\mu^{(5)}/M_b$
is displayed for the various levels of accuracy, ranging from one-loop to
four-loop evolution.
For illustration, $\mu^{(5)}$ is varied rather extremely, by almost two orders
of magnitude.
While the leading-order result exhibits a strong logarithmic behaviour, the
analysis is gradually getting more stable as we go to higher orders.
The four-loop curve is almost flat.
Besides the $\mu^{(5)}$ dependence of $\alpha_s^{(5)}(M_Z)$, also its absolute 
normalization is significantly affected by the higher orders.
At the central scale $\mu^{(5)}=M_b$, we again encounter an alternating
convergence behaviour.
We notice that the four-loop result is appreciably smaller than the three-loop
result, by almost 0.001.
This difference is comparable in size to the shift in the value of
$\alpha_s^{(5)}(M_Z)$ extracted from the measured $Z$-boson hadronic decay
width due to the inclusion of the known three-loop correction to this
observable \cite{che}.

As we have learned from Fig.~\ref{f3}, in higher orders, the actual value of
$\mu^{(n_f)}$ does not matter as long as it is comparable to the heavy-quark
mass.
In the context of Eq.~(\ref{msb}), the choice $\mu^{(n_f)}=\mu_h$ \cite{pdg} 
is particularly convenient, since it eliminates the RG logarithm $\ell_h$.
With this convention, we obtain from Eqs.~(\ref{con}), (\ref{alp}), and
(\ref{msb}) a simple relationship between $\Lambda^\prime=\Lambda^{(n_l)}$ and
$\Lambda=\Lambda^{(n_f)}$, viz
\begin{eqnarray}
\label{lam}
&&\beta_0^\prime\ln\frac{\Lambda^{\prime2}}{\Lambda^2}=
(\beta_0^\prime-\beta_0)l_h+(b_1^\prime-b_1)\ln l_h
-b_1^\prime\ln\frac{\beta_0^\prime}{\beta_0}
\nonumber\\
&&{}+\frac{1}{\beta_0l_h}\left[b_1(b_1^\prime-b_1)\ln l_h
+b_1^{\prime2}-b_1^2-b_2^\prime+b_2+c_2\right]
\nonumber\\
&&{}+\frac{1}{(\beta_0l_h)^2}\left\{-\frac{b_1^2}{2}(b_1^\prime-b_1)
\ln^2l_h+b_1[-b_1^\prime(b_1^\prime-b_1)
\right.\nonumber\\
&&{}+b_2^\prime-b_2-c_2]\ln l_h
+\frac{1}{2}(-b_1^{\prime3}-b_1^3-b_3^\prime+b_3)
\nonumber\\
&&{}+\left.\vphantom{\frac{b_1^2}{2}}
b_1^\prime(b_1^2+b_2^\prime-b_2-c_2)+c_3\right\},
\end{eqnarray}
where $l_h=\ln(\mu_h^2/\Lambda^2)$.
The ${\cal O}(1/l_h^2)$ term of Eq.~(\ref{lam}) represents a new result.
Leaving aside this term, Eq.~(\ref{lam}) disagrees with Eq.~(9.7) of 
Ref.~\cite{pdg}.
This disagreement may partly be traced to the fact that the latter equation is
written with the $c_2$ value obtained in Ref.~\cite{ber}, which differs from
the value listed in Eq.~(\ref{cms}).
Furthermore, in the same equation, the terms involving $\beta_2$ should be
divided by 4.
Equation~(\ref{lam}) represents a closed three-loop formula for
$\Lambda^{(n_l)}$ in terms of $\Lambda^{(n_f)}$ and $\mu_h$.
For consistency, it should be used in connection with the four-loop
expression~(\ref{alp}) for $\alpha_s^{(n_f)}(\mu)$ with the understanding that
the underlying flavour thresholds are fixed at $\mu^{(n_f)}=\mu_h$.
The inverse relation that gives $\Lambda^{(n_f)}$ as a function of
$\Lambda^{(n_l)}$ and $\mu_h$ emerges from Eq.~(\ref{lam}) via the 
substitutions $\Lambda\leftrightarrow\Lambda^\prime$;
$\beta_N\leftrightarrow\beta_N^\prime$ for $N=0,\ldots,3$; and $c_N\to-c_N$
for $N=2,3$.
The on-shell version of Eq.~(\ref{lam}), appropriate to the choice
$\mu^{(n_f)}=M_h$, is obtained by substituting
$l_h\to L_h=\ln(M_h^2/\Lambda^2)$ and $c_n\to C_N$ for $N=2,3$.
Analogously to the case of $\mu^{(n_f)}=\mu_h$, its inverse, which gives
$\Lambda^{(n_f)}$ in terms of $\Lambda^{(n_l)}$ and $M_h$, then follows
through the replacements $\Lambda\leftrightarrow\Lambda^\prime$;
$\beta_N\leftrightarrow\beta_N^\prime$ for $N=0,\ldots,3$; and $C_N\to-C_N$
for $N=2,3$.

In conclusion, we have extended the standard description of the strong
coupling constant in the $\overline{\rm MS}$ renormalization scheme to include
four-loop evolution and three-loop matching at the quark-flavour thresholds.
As a by-product of our analysis, we have settled a conflict in the literature
regarding the two-loop matching conditions \cite{ber,lar}.
These results will be indispensible in order to relate the QCD predictions for
different observables at next-to-next-to-next-to-leading order.
Meaningful estimates of such corrections already exist \cite{sam}.


\begin{thebibliography}{99}

\bibitem{pdg} Particle Data Group, R.M. Barnett {\it et al.},
Phys.\ Rev.\ D {\bf54}, 1 (1996).

\bibitem{gro} D.J. Gross and F. Wilczek,
Phys.\ Rev.\ Lett.\ {\bf30}, 1343 (1973); Phys.\ Rev.\ D {\bf8}, 3633 (1973);
H.D. Politzer, Phys.\ Rev.\ Lett.\ {\bf30}, 1346 (1973).

\bibitem{bol} C.G. Bollini and J.J. Giambiagi,
Phys.\ Lett.\ {\bf40B}, 566 (1972);
G. 't~Hooft and M. Veltman, Nucl.\ Phys.\ {\bf B44}, 189 (1972);
G. 't~Hooft, Nucl.\ Phys.\ {\bf B61}, 455 (1973).

\bibitem{jon} D.R.T. Jones, Nucl.\ Phys.\ {\bf B75}, 531 (1974);
W.E. Caswell, Phys.\ Rev.\ Lett.\ {\bf33}, 244 (1974);
\'E.Sh.\ Egoryan and O.V. Tarasov, Teor.\ Mat.\ Fiz.\ {\bf41}, 26 (1979)
[Theor.\ Math.\ Phys.\ {\bf41}, 863 (1979)].

\bibitem{tar} O.V. Tarasov, A.A. Vladimirov, and A.Yu.\ Zharkov,
Phys.\ Lett.\ {\bf93B}, 429 (1980);
S.A. Larin and J.A.M. Vermaseren, Phys.\ Lett.\ B {\bf303}, 334 (1993).

\bibitem{rit} T. van Ritbergen, J.A.M. Vermaseren, and S.A. Larin,
Phys.\ Lett.\ B {\bf400}, 379 (1997).

\bibitem{bar} W.A. Bardeen, A.J. Buras, D.W. Duke, and T. Muta,
Phys.\ Rev.\ D {\bf18}, 3998 (1978).

\bibitem{app} T. Appelquist and J. Carazzone,
Phys.\ Rev.\ D {\bf11}, 2856 (1975).

\bibitem{rod} G. Rodrigo and A. Santamaria,
Phys.\ Lett.\ B {\bf313}, 441 (1993).

\bibitem{ber} W. Wetzel,
Nucl.\ Phys.\ {\bf B196}, 259 (1982);
W. Bernreuther and W. Wetzel,
Nucl.\ Phys.\ {\bf B197}, 228 (1982);
W. Bernreuther,
Ann.\ Phys.\ {\bf151}, 127 (1983); Z. Phys.\ C {\bf20}, 331 (1983).

\bibitem{lar} S.A. Larin, T. van Ritbergen, and J.A.M. Vermaseren,
Nucl.\ Phys.\ {\bf B438}, 278 (1995).

\bibitem{fur} W. Furmanski and R. Petronzio, Z. Phys.\ C {\bf11}, 293 (1982).

\bibitem{abb} L.F. Abbott, Phys.\ Rev.\ Lett.\ {\bf44}, 1569 (1980);
E. Monsay and C. Rosenzweig, Phys.\ Rev.\ D {\bf23}, 1217 (1981);
W.A. Bardeen (private communication).

\bibitem{mar} W.J. Marciano, Phys.\ Rev.\ D {\bf29}, 580 (1984).

\bibitem{nog} P. Nogueira,
J. Comput.\ Phys.\ {\bf105}, 279 (1993).

\bibitem{ste} M. Steinhauser, Ph.D. thesis, Karlsruhe University
(Shaker Verlag, Aachen, 1996).

\bibitem{kni} B.A. Kniehl and M. Steinhauser,
Nucl.\ Phys.\ {\bf B454}, 485 (1995).

\bibitem{gra} N. Gray, D.J. Broadhurst, W. Grafe, and K. Schilcher,
Z. Phys.\ C {\bf48}, 673 (1990);
D.J. Broadhurst, N. Gray, and K. Schilcher, Z. Phys.\ C {\bf52}, 111 (1991).

\bibitem{che} K.G. Chetyrkin, J.H. K\"uhn, and A. Kwiatkowski,
Phys.\ Rep.\ {\bf277}, 189 (1996).

\bibitem{sam} M.A. Samuel, J. Ellis, and M. Karliner,
Phys.\ Rev.\ Lett.\ {\bf74}, 4380 (1995);
A.L. Kataev and V.V. Starshenko,
Mod.\ Phys.\ Lett.\ A {\bf10}, 235 (1995); Phys.\ Rev.\ D {\bf52}, 402 (1995);
P.A. R\c aczka and A. Szymacha,
Z. Phys.\ C {\bf70}, 125 (1996); Phys.\ Rev.\ D {\bf54}, 3073 (1996);
J. Ellis, E. Gardi, M. Karliner, and M.A. Samuel,
Phys.\ Lett.\ B {\bf366}, 268 (1996); Phys.\ Rev.\ D {\bf54}, 6986 (1996);
K.G. Chetyrkin, B.A. Kniehl, and A. Sirlin,
Report Nos.\ MPI/PhT/97--010, NYU--TH--97/03/01, and hep--ph/9703226
(February 1997), Phys.\ Lett.\ B (in press);
S. Groote, J.G. K\"orner, A.A. Pivovarov, and K. Schilcher,
Report Nos.\ MZ--TH/97--09 and hep--ph/9703208 (March 1997);
S. Groote, J.G. K\"orner, and A.A. Pivovarov,
Report Nos.\ MZ--TH--97--16 and hep-ph/9704396 (April 1997).

\end{thebibliography}
\end{document}